\documentclass[12pt]{article}
\usepackage{authblk}
\usepackage[utf8]{inputenc}
\usepackage[margin=0.8in]{geometry}
\usepackage{blindtext, xfrac}
\usepackage[english]{babel}
\usepackage[T1]{fontenc}
\usepackage{titling}
\usepackage{amsmath}
\usepackage{amssymb}
\usepackage{cite}
\usepackage{graphicx}
\usepackage{booktabs}
\usepackage{physics}
\usepackage{nameref}
\usepackage{ bbold }

\usepackage{textgreek}
\usepackage{hyperref}
\usepackage{gensymb}
\usepackage{float}
\usepackage[usenames,dvipsnames]{color}
\usepackage{svg}
\usepackage{placeins}
\usepackage{authblk}

\definecolor{greeny}{rgb}{0.2,0.8,0.5}

\title{Heavy Particle Decay Renders $N_{eff}$ an Ineffective Probe for Dark Sectors}
\title{How Effective is $N_{eff}$ at Discovering Dark Radiation in a Cosmology with Heavy Particle Decay?}

\author{Katarina Bleau$^1$, Joseph Bramante$^{1,2}$, and Christopher Cappiello$^{1,2}$
\\
$^1$The Arthur B. McDonald Canadian Astroparticle Physics Research Institute, \protect\\ Department of Physics, Engineering Physics, and Astronomy, \protect\\ Queen's University, Kingston, Ontario, K7L 2S8, Canada
\\
$^2$Perimeter Institute for Theoretical Physics, Waterloo, ON N2J 2W9, Canada}

\date{\today}

\begin{document}
\maketitle

\begin{abstract}

Any light relic which was in thermal equilibrium with the Standard Model before it freezes out results in a shift in the effective number of neutrino species, $N_{eff}$. This quantity is being measured with increasing precision, and planned experiments would seemingly rule out light particles beyond the Standard Model, even for rather high temperature light particle freeze out. Here we explore how these bounds are loosened if the energy density of the light particles is diluted with respect to that of Standard Model radiation, which can happen if a heavy particle decays into the Standard Model bath after the light particle freezes out. After calculating how heavy state decays alter $N_{eff}$ for light particles beyond the Standard Model,  we focus in particular on the case that the heavy decaying particle is a gravitino, and use current bounds on $N_{eff}$ to place constraints on the gravitino mass and the branching ratio into light particles for different values of the reheating temperature of the Universe.

\end{abstract}

\section{Introduction}

Astrophysical observations such as galactic rotation curves \cite{rotation_curves, Rubin:1980zd}, gravitational lensing \cite{bullet_cluster}, as well as large-scale structure and the CMB power spectrum \cite{CMB_power_spectra} provide compelling evidence for the existence of dark matter. Many proposed dark matter models contain particles much lighter than the typical WIMP scale, either as the dark matter itself or as a new mediator, ranging from MeV mass scales down to ultralight dark photons and axions~\cite{Dine:1982ah,Preskill:1982cy,Holdom:1985ag,Masso:1995tw,Hu:2000ke,McDonald:2001vt,Boehm:2003hm,Feng:2008ya,Feng:2008mu,Hooper:2008im,Kaplinghat:2013yxa,Boddy:2014yra,Knapen:2017xzo}.
The existence of light particles beyond the Standard Model would affect the total energy density in radiation of the Universe, which can be parameterized by the effective number of neutrino species $N_{eff}$. Indeed, any light relic which was in thermal equilibrium with neutrinos at some time before freezing out will result in a change in $N_{eff}$, denoted $\Delta N_{eff}$, which is reviewed and derived in Appendix \ref{appendix}. Measurements of $N_{eff}$ constrain the mass of such relics to be greater than $\sim$a few MeV~\cite{Boehm:2012gr,Boehm:2013jpa,Sabti:2019mhn,Chu:2022xuh}.

$N_{eff}$ will be measured more precisely than ever by upcoming cosmic microwave background experiments (CMB-S4) \cite{CMBS4} and the Simons Observatory \cite{SimonsObservatory}, and the resulting sensitivity to this parameter could seemingly rule out or discover most candidate light particles beyond the Standard Model. However, resulting bounds on light particles would be loosened if the energy density of these light particles is diluted with respect to that of Standard Model radiation. This can happen if a heavy particle decays into the Standard Model bath after the light particle has frozen out. A heavy state decaying in the early Universe is a feature of many theories. For example, it could be a field associated with the Affleck-Dine mechanism of baryogenesis \cite{Affleck:1984fy}, a string modulus \cite{Banks:1993en,deCarlos:1993wie,Moroi:1999zb}, a heavy right-handed neutrino which generates the neutrino mass via the seesaw mechanism \cite{Randall2016}, or a gravitino \cite{Weinberg:1982zq}, the latter of which we consider here as a concrete example.

The impact of a gravitino or other heavy decaying state on the cosmology of dark sectors has been considered previously, especially in the context of the decay of heavy fields diluting dark matter or lepton/baryon number in the Standard Model \cite{Banks:1993en,deCarlos:1993wie,Moroi:1999zb,Affleck:1984fy,Dine:1995kz}. However, the impact of a heavy decaying state on $\Delta N_{eff}$ has been less explored -- although for recent work on the dilution of radiation in a dark sector in the context light dark matter abundances, see \cite{Randall2016}.

Consequently, the aim of this work is to carefully calculate the contribution to the effective number of neutrino species $\Delta N_{eff}$ in a scenario with both a light particle beyond the Standard Model and a heavy state decay. In the case where this heavy state is a gravitino, we constrain the gravitino mass and branching ratio. The following will be divided into three main sections. In Sections \ref{sec:SM} and \ref{sec:SM_LP}, we will calculate $\Delta N_{eff}$ for a heavy particle decaying into only Standard Model radiation, and into both the Standard Model and the light particle, respectively. In Section \ref{sec:gravitino}, we will numerically calculate $\Delta N_{eff}$ for a range of parameters in the case where the heavy particle is a gravitino, allowing us to place bounds on the gravitino mass and the branching ratio for decay into the light particle at different reheating temperatures. In Section \ref{sec:conc} we conclude. Appendix \ref{appendix} reviews the derivation of $\Delta N_{eff}$ in cosmologies without heavy state decay.

\section{$N_{eff}$ With Heavy State Decay into the Standard Model}
\label{sec:SM}

The effect of extra light degrees of freedom in equilibrium with the Standard Model thermal bath in the early universe has been well studied - see $e.g.$ \cite{KolbTurner,Planck2018}. We review the standard treatment in Appendix \ref{appendix}, where the contribution to the extra number of effective relativistic degrees of freedom is derived to be
\begin{equation}
    \Delta N_{eff} = 
\begin{cases}
    \frac{4g}{7} \left(\frac{43/4}{g_*(T_F)} \right)^{4/3} & \text{for a boson}, \\
    \frac{g}{2} \left(\frac{43/4}{g_*(T_F)} \right)^{4/3} & \text{for a fermion},
\end{cases}
\label{eq:N_eff}
\end{equation}
where $g_*$ is the number of relativistic degrees of freedom in the Standard Model at the time that the (effectively massless) boson or fermion decoupled from the thermal bath and $g$ is the number of degrees of freedom of the new light particle.

We now turn to the effect a heavy decaying field would have on $\Delta N_{eff}$. At first, in keeping with the standard scenario, we will consider a light particle (LP) that has decoupled from the Standard Model bath in the early Universe. Then, a heavy particle (HP) decays into Standard Model (SM) radiation before neutrino decoupling, increasing the entropy density of the latter with respect to the light particle. This entropy dump will result in a dilution of $\Delta N_{eff}$. First, we will calculate $\Delta N_{eff}$ in terms of a cosmological abundance dilution factor \cite{Bramante2017,Randall2016}
\begin{equation}
    \zeta \equiv \frac{a_{before}^3 s_{before}}{a_{after}^3 s_{after}},
\label{eq:zeta}
\end{equation}
which is a ratio composed of the comoving entropy density of the Standard Model bath before and after the decay of the heavy state, where $a$ is the scale factor of the Universe and $s$ is the entropy density of the Standard Model bath \cite{Bramante2017}. Next, we will lay out two methods for computing $\zeta$, which provide an exact solution and an approximate solution respectively: numerically solving the Friedmann equations, and assuming that the decay happens suddenly at a time where the heavy state dominates the energy density of the Universe. 

To calculate $\Delta N_{eff}$, we start by relating the comoving entropy density of the Standard Model neutrinos before and after the heavy state decay, recalling $s = \frac{2\pi^2}{45}g_*T^3$ for relativistic particles:
\begin{equation}
    a_{before}^3 g_*^{before} T_{\nu, before}^3 = \zeta a_{after}^3 g_*^{after} T_{\nu, after}^3,
\label{eq:diluted_s_cons}
\end{equation}
where $g_*$ is the number of spin states defined by Eq. \ref{eq:g*} in Appendix \ref{appendix} and $T$ is the temperature. Since the comoving entropy density of the light particle is conserved, and we assume the number of light particle degrees of freedom remains fixed, 
\begin{equation}
    a_{before}^3 T_{LP, before}^3 = a_{after}^3 T_{LP, after}^3.
\label{eq:LP_s_cons}
\end{equation}
Dividing Eq. \ref{eq:diluted_s_cons} by Eq. \ref{eq:LP_s_cons},
\begin{equation}
    \frac{g_*^{before} T_{\nu, before}^3}{T_{LP, before}^3} = \frac{\zeta g_*^{after} T_{\nu, after}^3}{T_{LP, after}^3},
\end{equation}
which can be rearranged to solve for the temperature of light particles with respect to that of neutrinos after the decay:
\begin{equation}
    \frac{T_{LP, after}^3}{T_{\nu, after}^3} = \zeta \frac{g_*^{after}}{g_*^{before}} \frac{T_{LP, before}^3}{T_{\nu, before}^3}.
\label{eq:T_frac}
\end{equation}
Here, $g_*^{before} = g_*^{after}$ because we assume the number of Standard Model spin states/degrees of freedom remain fixed during the heavy state decay. Furthermore, the temperature fraction before the decay can be replaced by Eq. \ref{eq:T_LP_nu}, whose derivation can be found in Appendix \ref{appendix}. Therefore, Eq. \ref{eq:T_frac} can be rewritten as
\begin{equation}
    \frac{T_{LP, after}^3}{T_{\nu, after}^3} = \zeta \frac{43/4}{g_*(T_F)}.
\label{eq:T_frac_reduced}
\end{equation}
In Appendix \ref{appendix}, we derive $\Delta N_{eff}$ for any light particle which was in thermal equilibrium with the Standard Model at some time before freezing out. In terms of the temperature fraction, it can be written as (suppressing ``after'' subscripts for simplicity)
\begin{equation}
    \Delta N_{eff} = 
\begin{cases}
    \frac{4g}{7} \left(\frac{T_{LP}}{T_{\nu}} \right)^{4} & \text{for a boson}, \\
    \frac{g}{2} \left(\frac{T_{LP}}{T_{\nu}} \right)^{4} & \text{for a fermion.}
\end{cases}
\label{eq:N_eff_T}
\end{equation}
Substituting Eq. \ref{eq:T_frac_reduced} into Eq. \ref{eq:N_eff_T}, we find $\Delta N_{eff}$ to be
\begin{equation}
    \Delta N_{eff} = 
\begin{cases}
    \frac{4g}{7} \left(\zeta \frac{43/4}{g_*(T_F)} \right)^{4/3} & \text{for a boson}, \\
    \frac{g}{2} \left(\zeta \frac{43/4}{g_*(T_F)} \right)^{4/3} & \text{for a fermion}
\end{cases}
\end{equation}
with the inclusion of a heavy particle decay. Therefore, a heavy state decaying into the Standard Model bath before neutrino decoupling and after a light particle has frozen out decreases $\Delta N_{eff}$ by a factor of $\zeta^{4/3}$. When the heavy particle decays, it dumps entropy into the Standard Model, but not into the light particle bath since it has already frozen out. Consequently, the energy density of the light particles becomes less significant compared to the energy density of the Standard Model, which results in a smaller $\Delta N_{eff}$.

\begin{figure}[h!]
    \centering
    \includegraphics[scale=0.8]{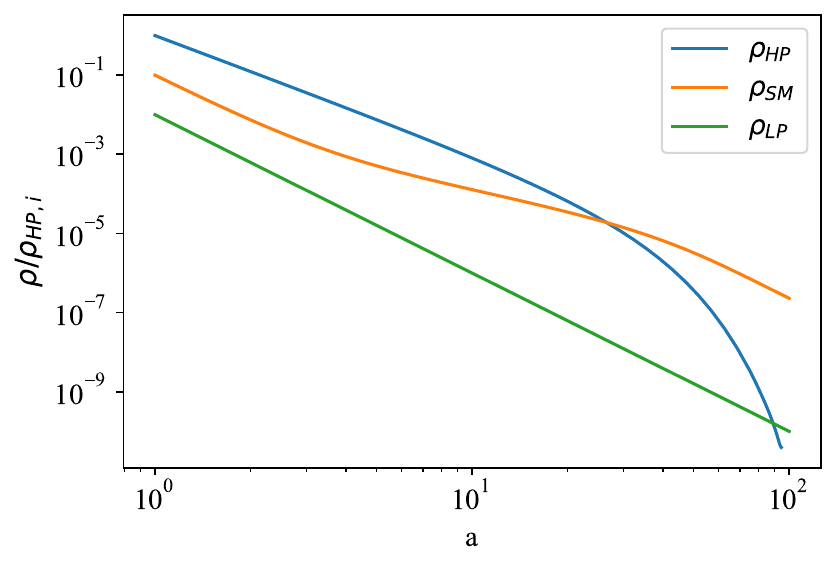}
    \caption{Energy density scaled by the initial energy density of the heavy particle as a function of the scale factor of the universe. The exact solutions for the heavy particle, the Standard Model particles and the light particles are shown in blue, orange and green respectively. The initial conditions for the system of ODEs are $\rho_{HP, i}$ = 100 GeV$^4$, $\rho_{SM, i}$ = 10 GeV$^4$ and $\rho_{LP, i}$ = 1 GeV$^4$. This roughly corresponds to the case of the heavy field decoupling at $a \sim 10^{-3}$. The decay rate of the heavy particle is $H_i/100$, where $H_i$ is determined using the initial SM, HP, LP energy densities. Since the energy density of the Standard Model particles increases after the decay, the energy density of the light particles gets diluted.}
    \label{fig:LP_dilution}
\end{figure}

The dilution factor $\zeta$ can be calculated by rewriting the entropy density of the Standard Model bath $s$ in terms of its energy density $\rho_{SM}$. By recalling the equations for the energy density in radiation $\rho_{SM} = \frac{\pi^2}{30}g_* T^4$ and the entropy density $s = \frac{2\pi^2}{45} g_* T^3$ for relativistic particles, we find
\begin{equation}
    s = \frac{4}{3} \left( \frac{\pi^2}{30} g_* \right)^{1/4} \rho_{SM}^{3/4}.
\label{eq:s_vs_rho}
\end{equation}
Inserting this expression into Eq. \ref{eq:zeta} allows us to determine $\zeta$ in terms of the energy density of Standard Model radiation. It is possible to calculate $\rho_{SM}$ exactly by numerically solving the Friedmann equations or approximately by assuming that the heavy particle dominates the energy density of the Universe before decaying suddenly \cite{Randall2016}.

For an exact solution, we start with the following system of equations:
\begin{equation}
    \Dot{\rho_{HP}} + 3H\rho_{HP} = -\Gamma \rho_{HP},
\end{equation}
\begin{equation}
    \Dot{\rho_{SM}} + 4H\rho_{SM} = \Gamma \rho_{HP},
\label{eq:rho_SM}
\end{equation}
\begin{equation}
    \Dot{\rho_{LP}} + 4H\rho_{LP} = 0,
\label{eq:rho_LP}
\end{equation}
where
\begin{equation}
    H^2 = \frac{8\pi}{3M_{Pl}^2}(\rho_{HP} + \rho_{SM} + \rho_{LP})
\end{equation}
and $\Gamma$ is the decay rate of the heavy particle. The derivation of Eq. \ref{eq:rho_SM} and Eq. \ref{eq:rho_LP} is only valid when $g_*$ is constant \cite{KolbTurner}. It is useful to rewrite these equations to solve for the energy densities as functions of the scale factor $a$ instead of the time $t$:
\begin{equation}
    aH\rho_{HP}' + 3H\rho_{HP} = -\Gamma \rho_{HP},
\end{equation}
\begin{equation}
    aH\rho_{SM}' + 4H\rho_{SM} = \Gamma \rho_{HP},
\end{equation}
\begin{equation}
    aH\rho_{LP}' + 4H\rho_{LP} = 0,
\end{equation}
where a prime denotes a derivative with respect to $a$. This system of ODEs can be solved for $\rho_{HP}(a)$, $\rho_{SM}(a)$ and $\rho_{LP}(a)$ given initial conditions and a decay rate $\Gamma$. Figure \ref{fig:LP_dilution} shows the exact solutions for $\rho_{HP}(a)$, $\rho_{SM}(a)$ and $\rho_{LP}(a)$, which allows us to observe the dilution of the light particle bath with respect to the Standard Model bath. 

\begin{figure}[h!]
    \centering
    \includegraphics[width=0.65\columnwidth]{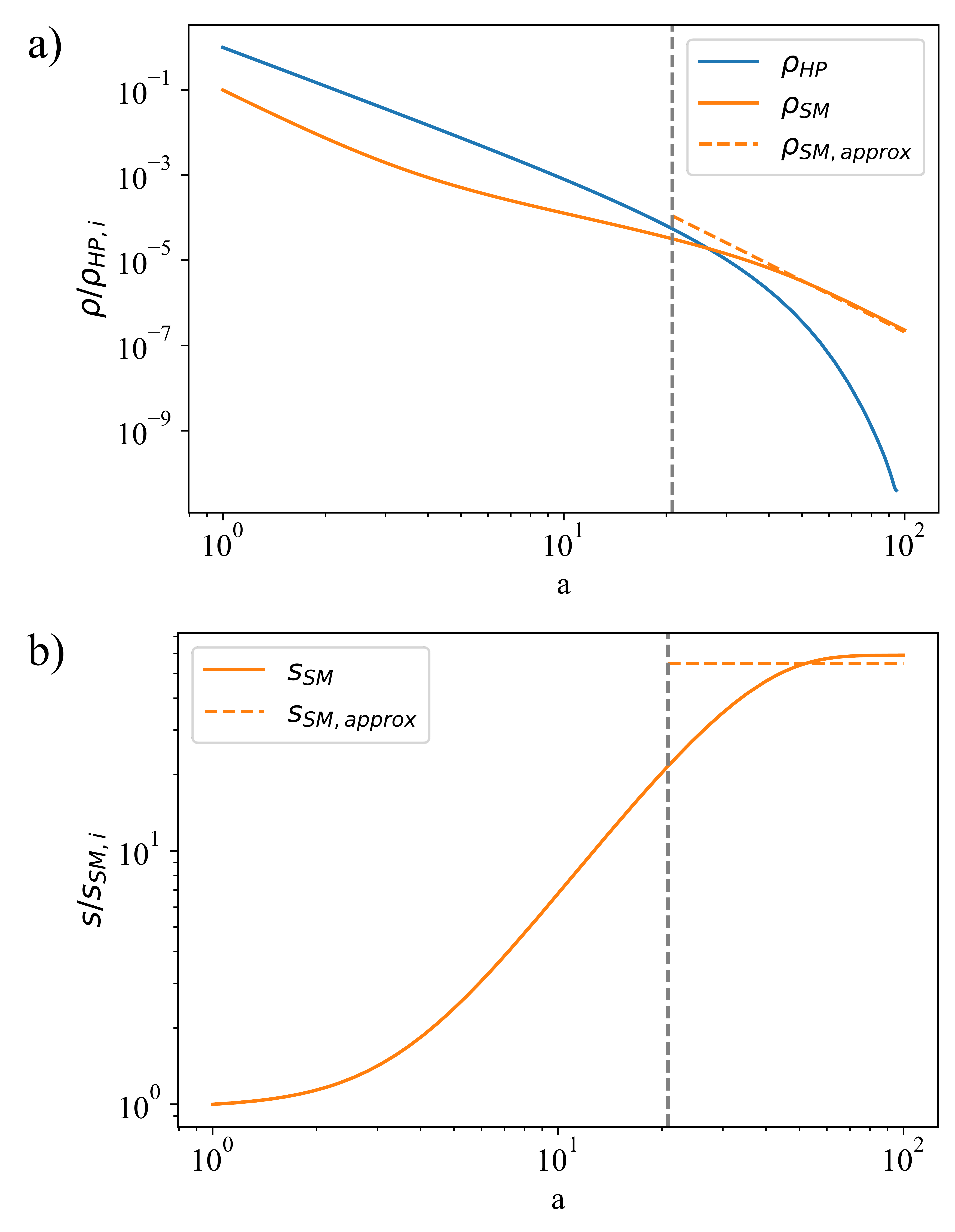}
    \caption{a) Energy density scaled by the initial energy density of the heavy particle\textcolor{red}{,} as a function of the scale factor of the universe. The exact solutions for the heavy particle and for the Standard Model particles are shown in blue and orange respectively, and the sudden decay approximation for $\nu$ = 1 is represented by a dashed orange line. The initial conditions for the system of ODEs are $\rho_{HP, i}$ = 100 GeV$^4$, $\rho_{SM, i}$ = 10 GeV$^4$ and $\rho_{LP, i}$ = 1 GeV$^4$. The decay rate of the heavy particle is $H_i/100$. The dashed vertical line is drawn at $a_*$. b) Comoving entropy density scaled by the initial comoving entropy density of the Standard Model as a function of the scale factor of the universe. The exact solution for the Standard Model particles is represented by a solid line and the sudden decay approximation for $\nu$ = 1 is represented by a dashed line. The dashed vertical line is drawn at $a_*$. Setting $\nu$ = 1 causes a slight underestimation of the energy density of the Standard Model after the decay, which could be explained by fact that some of heavy state decays after $a_*$ and contributes to the energy density of the Standard Model.}
    \label{fig:exact_vs_approx_SM}
\end{figure}

For an approximate solution, we begin by assuming that the heavy particle dominates the energy density of the Universe before decaying (at $a_i$), in order that
\begin{equation}
    H^2 = \frac{8\pi}{3M_{Pl}^2} \rho_{HP,i}.
\end{equation}
Then, we assume that the decay happens suddenly such that $H = \Gamma/\nu$ at the time of decay ($a_*$), where $\nu$ is a fit parameter that can be determined by comparing the approximate solution to the exact solution. Therefore, at $a_*$, the energy density of the heavy particle is given by
\begin{equation}
    \rho_{HP,*} = \frac{3 M_{Pl}^2 \Gamma^2}{8\pi\nu^2}.
\label{eq:rho_HP_approx}
\end{equation}
Since $\rho_{HP}$ evolves like matter,
\begin{equation}
    \frac{\rho_{HP,i}}{\rho_{HP,*}} = \left( \frac{a_*}{a_i} \right)^3.
\end{equation}
Setting $a_i = 1$, we find that the scale factor at the time of decay is
\begin{equation}
    a_* = \left( \frac{8\pi\nu^2 \rho_{HP,i}}{3 M_{Pl}^2 \Gamma^2} \right)^{1/3}.
\end{equation}
Since $\rho_{SM}$ evolves like radiation,
\begin{equation}
    \frac{\rho_{SM}}{\rho_{SM,*}} = \left( \frac{a_*}{a} \right)^4,
\label{eq:rhoSM_evol}
\end{equation}
where $\rho_{SM,*} \simeq \rho_{HP,*}$ in the sudden decay approximation. Therefore, the energy of the Standard model after the heavy particle decays is
\begin{equation}
    \rho_{SM} = \rho_{HP,*} \left( \frac{a_*}{a} \right)^4,
\end{equation}
where $\rho_{HP,*}$ is given by Eq. \ref{eq:rho_HP_approx}.
From Figure \ref{fig:exact_vs_approx_SM}a, it can be shown that $\nu$ = 1 gives a good approximation for $\rho_{SM}$ compared to the exact solution, as opposed to previous work which found $\nu$ = 3 \cite{Randall2016}. Figure \ref{fig:exact_vs_approx_SM} shows the exact solution and the approximate solution for both the energy density and the comoving entropy density. We can observe that setting $\nu$ = 1 causes a slight underestimation of the energy density of the Standard Model after the decay. This could be explained by fact that some of the heavy state decays after $a_*$ and contributes to the energy density of the Standard Model, which isn't taken into account by the sudden decay approximation which assumes that all of the heavy state decays at $a_*$.

\section{$N_{eff}$ With Heavy State Decay into New Light Particle and the Standard Model}
\label{sec:SM_LP}

Next, we consider the case where a heavy particle (HP) decays into both Standard Model (SM) radiation and a decoupled light particle (LP) before neutrino decoupling, changing the relative entropy density of the Standard Model bath and the light particles. This entropy dump may result in either an increase or a decrease in $\Delta N_{eff}$ depending on the fraction of the heavy particles which decay into light particles. We will start by calculating $\Delta N_{eff}$ in this scenario and comparing it to $\Delta N_{eff}$ in the case where the heavy particle decays only into the Standard Model. Then, we will modify the exact and approximate methods for computing the dilution factor from the previous section to account for the decay into the light particle.

Similarly to the previous section, we write the comoving entropy density of both the Standard Model neutrinos and the light particles before the heavy state decay as following
\begin{equation}
    a_{before}^3 g_*^{SM, before} T_{\nu, before}^3 = \zeta_{SM} a_{after}^3 g_*^{SM, after} T_{\nu, after}^3,
\label{eq:SM_s_cons_2}
\end{equation}
\begin{equation}
    a_{before}^3 g_*^{LP, before} T_{LP, before}^3 = \zeta_{LP} a_{after}^3 g_*^{LP, after} T_{LP, after}^3.
\label{eq:LP_s_cons_2}
\end{equation}
We define an enhancement factor
\begin{equation}
    \epsilon_{LP} \equiv \frac{1}{\zeta_{LP}},
\label{eq:epsilon}
\end{equation}
that parameterizes the growth of the comoving entropy density of the light particles after the heavy state decays. Then, we divide Eq. \ref{eq:SM_s_cons_2} by Eq. \ref{eq:LP_s_cons_2} and rearrange to solve for the temperature of light particles with respect to that of neutrinos after the decay:
\begin{equation}
    \frac{T_{LP, after}^3}{T_{\nu, after}^3} = \zeta_{SM} \epsilon_{LP} \frac{g_*^{SM, after}}{g_*^{SM, before}} \frac{g_*^{LP, before}}{g_*^{LP, after}} \frac{T_{LP, before}^3}{T_{\nu, before}^3}.
\end{equation}
Since $g_*^{before} = g_*^{after}$ for both the Standard model and the light particle, and Eq. \ref{eq:T_LP_nu} still holds, the temperature fraction can be simplified to
\begin{equation}
    \frac{T_{LP, after}^3}{T_{\nu, after}^3} = \zeta_{SM} \epsilon_{LP} \frac{43/4}{g_*^{SM}(T_F)}.
\label{eq:T_frac_2}
\end{equation}
Substituting Eq. \ref{eq:T_frac_2} into Eq. \ref{eq:N_eff_T}, we find
\begin{equation}
    \Delta N_{eff} = 
\begin{cases}
    \frac{4g}{7} \left(\zeta_{SM} \epsilon_{LP} \frac{43/4}{g_*^{SM}(T_F)} \right)^{4/3} & \text{for a boson}, \\
    \frac{g}{2} \left(\zeta_{SM} \epsilon_{LP} \frac{43/4}{g_*^{SM}(T_F)} \right)^{4/3} & \text{for a fermion}.
\end{cases}
\label{eq:DeltaNeff}
\end{equation}
Therefore, a heavy state decaying into both the Standard Model bath and light particles before neutrino decoupling and after the light particle has frozen out causes $\Delta N_{eff}$ to be multiplied by a factor of $\left( \zeta_{SM} \epsilon_{LP} \right)^{4/3}$. In the case where the heavy state decays into only the Standard Model, $\epsilon$ = 1 and we recover the result from the previous section. If the heavy particle decay causes the entropy of the Standard Model bath to increase by a larger proportional factor than the entropy of the light particle, $\Delta N_{eff}$ is smaller than in the case without heavy particle decay.

The dilution factor $\zeta_{SM}$ and the enhancement factor $\epsilon_{LP}$ can be calculated from Eq. \ref{eq:s_vs_rho} the same way as in the previous section, where the entropy density of the Standard Model and that of the light particles depend on the energy density of the Standard Model bath $\rho_{SM}$ and the light particle bath $\rho_{LP}$ respectively. We can then calculate these energy densities by generalizing the exact and approximate solutions from the previous section.

\begin{figure}[h!]
    \centering
    \includegraphics[width=0.65\columnwidth]{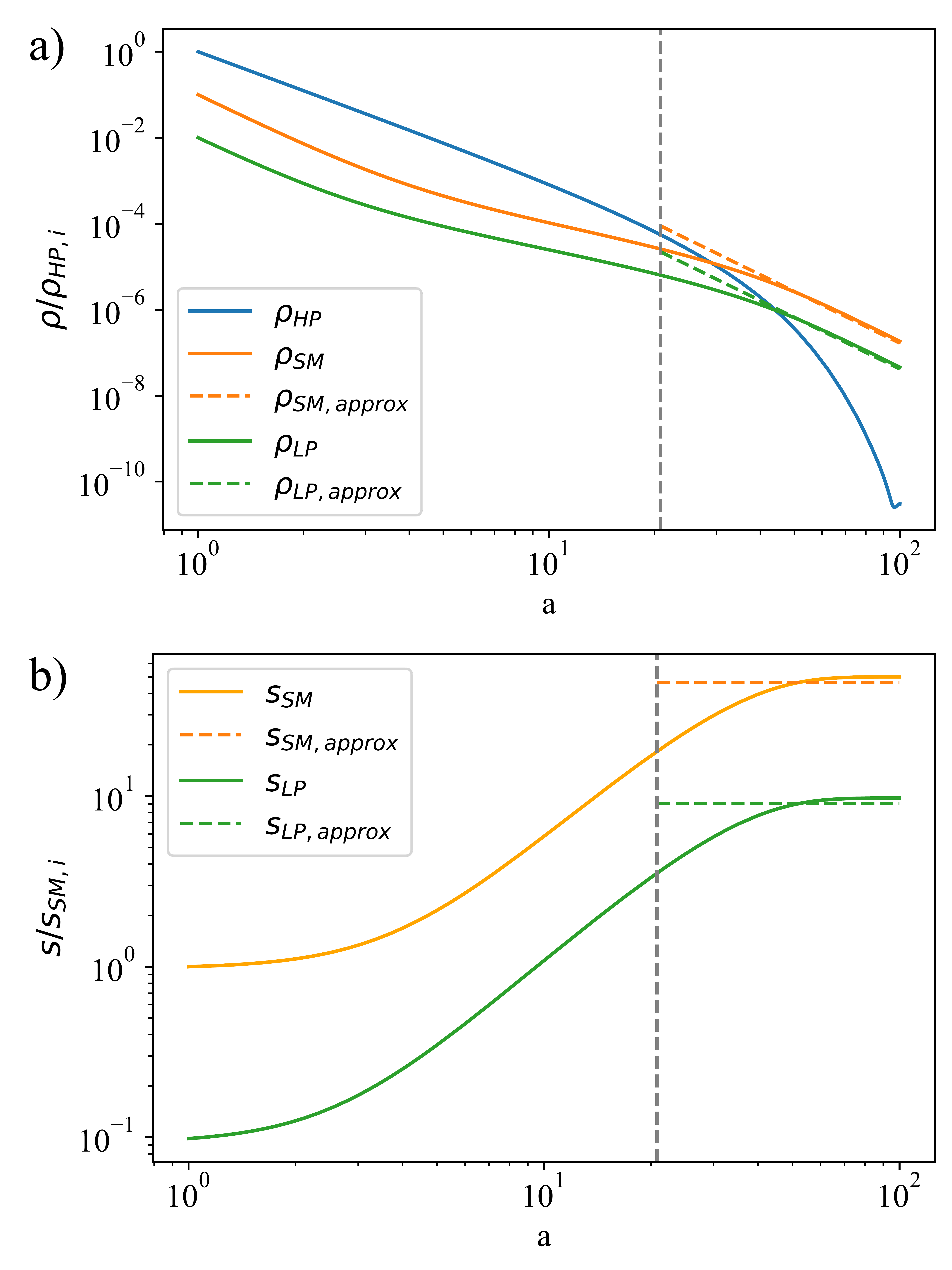}
    \caption{a) Energy density scaled by the initial energy density of the heavy particle as a function of the scale factor of the universe. The exact solutions for the heavy particle, for the Standard Model particles and for the nonstandard light particle are shown in blue, orange and green respectively. The sudden decay approximation for $\nu$ = 1 is represented by the orange and green dashed lines. The initial conditions for the system of ODEs are $\rho_{HP, i}$ = 100 GeV$^4$, $\rho_{SM, i}$ = 10 GeV$^4$ and $\rho_{LP, i}$ = 1 GeV$^4$. The decay rate of the heavy particle is $H_i/100$ and the branching ratio for the decay into the light particle is 0.2. b) Comoving entropy density scaled by the initial comoving entropy density of the Standard Model as a function of the scale factor of the universe. The exact solutions for the Standard Model particles and for the light particle are shown in orange and in green respectively. The sudden decay approximation for $\nu$ = 1 is represented by the orange and green dashed lines.}
    \label{fig:exact_vs_approx_SMLP}
\end{figure}

For the exact solution, we split the decay rate of the heavy particle into two terms, defining $\Gamma = \Gamma_{SM} + \Gamma_{LP}$, where $\Gamma_{SM}/\Gamma$ and $\Gamma_{LP}/\Gamma$ are the branching ratios for the decay into the Standard Model and into the light particle respectively. It is then possible to solve the system of equations from the previous section, this time accounting for the possibility of decay into light particles, for a given $\Gamma_{LP}/\Gamma$:
\begin{equation}
    aH\rho_{HP}' + 3H\rho_{HP} = -\Gamma \rho_{HP},
\end{equation}
\begin{equation}
    aH\rho_{SM}' + 4H\rho_{SM} = \Gamma \left( 1 - \frac{\Gamma_{LP}}{\Gamma} \right) \rho_{HP},
\end{equation}
\begin{equation}
    aH\rho_{LP}' + 4H\rho_{LP} = \Gamma \frac{\Gamma_{LP}}{\Gamma} \rho_{HP},
\end{equation}
where
\begin{equation}
    H^2 = \frac{8\pi}{3M_{Pl}^2}(\rho_{HP} + \rho_{SM} + \rho_{LP}).
\end{equation}

For the approximate solution, the derivation from the previous section stays true until Eq. \ref{eq:rhoSM_evol}, which holds for both the Standard Model radiation and the light particle. To account for the heavy particle decaying into both sectors in the sudden decay approximation, we set $\rho_{SM,*} = \left(1 - \frac{\Gamma_{LP}}{\Gamma} \right) \rho_{HP,*}$ and $\rho_{LP,*} = \frac{\Gamma_{LP}}{\Gamma} \rho_{HP,*}$ for a given $\Gamma_{LP}/\Gamma$. Therefore,
\begin{equation}
    \rho_{SM} = \left(1 - \frac{\Gamma_{LP}}{\Gamma} \right) \rho_{HP,*} \left( \frac{a_*}{a} \right)^4
\end{equation}
and
\begin{equation}
    \rho_{LP} = \frac{\Gamma_{LP}}{\Gamma} \rho_{HP,*} \left( \frac{a_*}{a} \right)^4,
\end{equation}
where $\rho_{HP,*}$ is given by Eq. \ref{eq:rho_HP_approx}. Figure \ref{fig:exact_vs_approx_SMLP} shows the exact solution and the approximate solution for both the energy density and the comoving entropy density.

\section{Gravitino Decay}
\label{sec:gravitino}

\begin{figure}[th!]
    \centering
    \includegraphics[width=0.78\columnwidth]{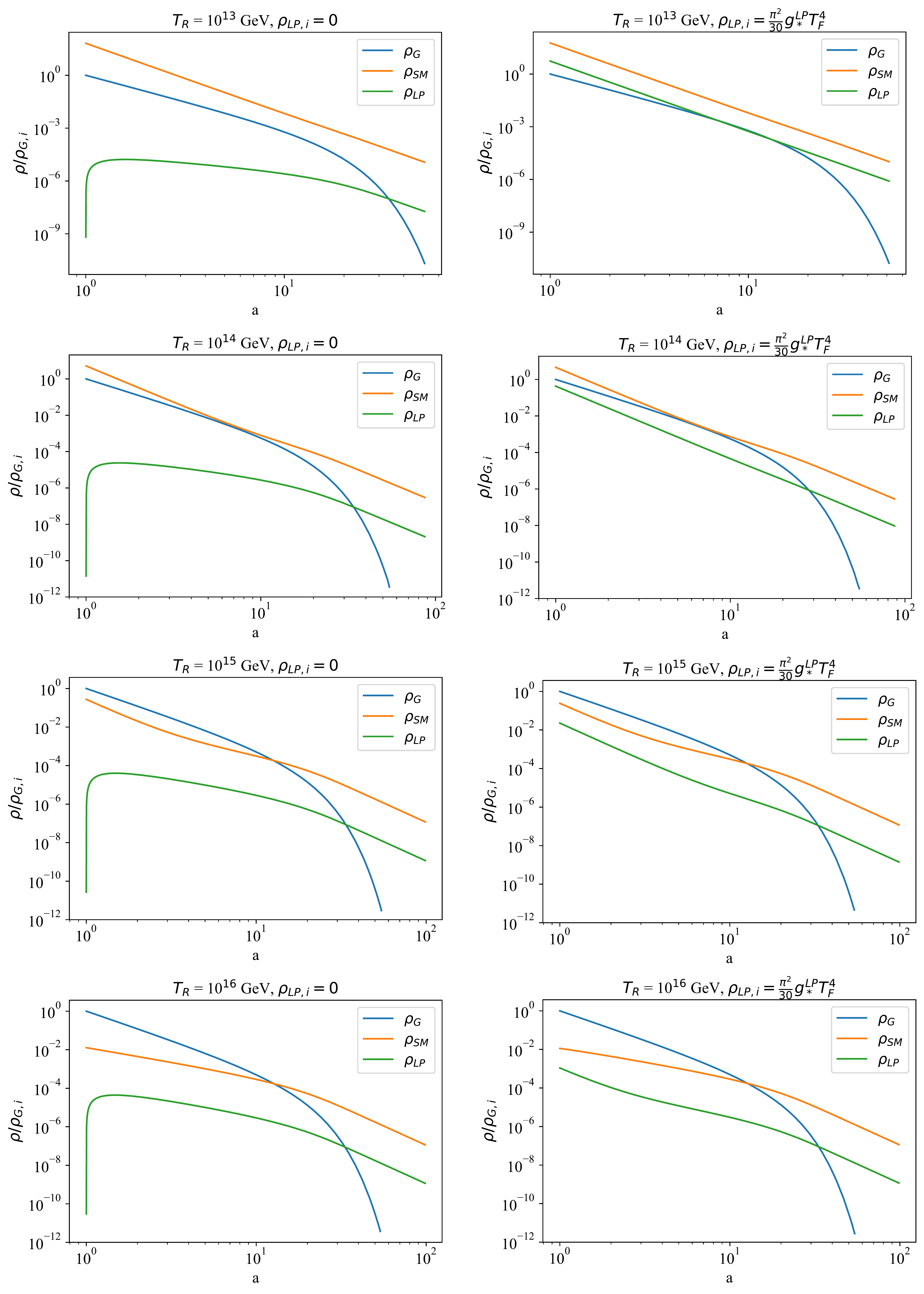}
    \caption{Energy density vs.~scale factor $a$ of the universe, where densities are scaled by the initial energy density of the gravitino for $m_G = 2 \times 10^4$ GeV,  $\Gamma_{LP}/\Gamma$ = 0.01, and  $\rho_{LP,i} = 0$ in the left column and $\rho_{LP,i} = \frac{\pi}{30}g_*^{LP}T_{F}^4$ in the right column. The reheating temperature varies between $T_R = 10^{13}-10^{16}$ GeV as indicated. The gravitino, the Standard Model and the light particle are represented by the blue line, the orange line and the green line respectively. At low reheating temperatures, the energy density of the gravitino is always smaller than that of the Standard Model, so the decay doesn't result in significant dilution. At high reheating temperatures, the energy density of the gravitino is bigger in comparison to that of the Standard Model just before the decay, resulting in a greater increase in the Standard Model energy density after the decay.}
    \label{fig:rhovsa}
\end{figure}

We now apply the formalism described above to a specific example, that of a gravitino decaying into the Standard Model bath and a decoupled light particle, where the light particle is a scalar or pseudoscalar boson. The gravitino is the spin-3/2 superpartner of the spin-2 graviton (see Refs.~\cite{Martin:1997ns,Nastase:2011aa} for reviews). Prior to supersymmetry breaking, it is massless, but through the breaking of supersymmetry it obtains a mass related to the supersymmetry breaking scale. In Ref.~\cite{Pagels:1981ke}, it was argued that if the gravitino is stable---as would be the case, for example, if it were the lightest supersymmetric particle in a model that respects R-parity---its mass must be $\lesssim$1 keV, or else its density would exceed the observed mass density of the universe. However, in R-parity violating models, or simply models where the gravitino is not the lightest supersymmetric particle, it may not be stable. Ref.~\cite{Weinberg:1982zq} showed that if the gravitino decays early enough, it can be prevented from dominating the present-day mass density of the universe or spoiling big bang nucleosynthesis. The latter requirement in particular leads to a lower bound on the heavy gravitino mass of $\sim$10 TeV. In this Section, we use $\Delta N_{eff}$ measurements to set bounds on the (heavy) gravitino mass and branching ratio for decay into the light particle, for different values of the reheating temperature.

\begin{figure}[h!]
    \centering
    \includegraphics[width=0.87\columnwidth]{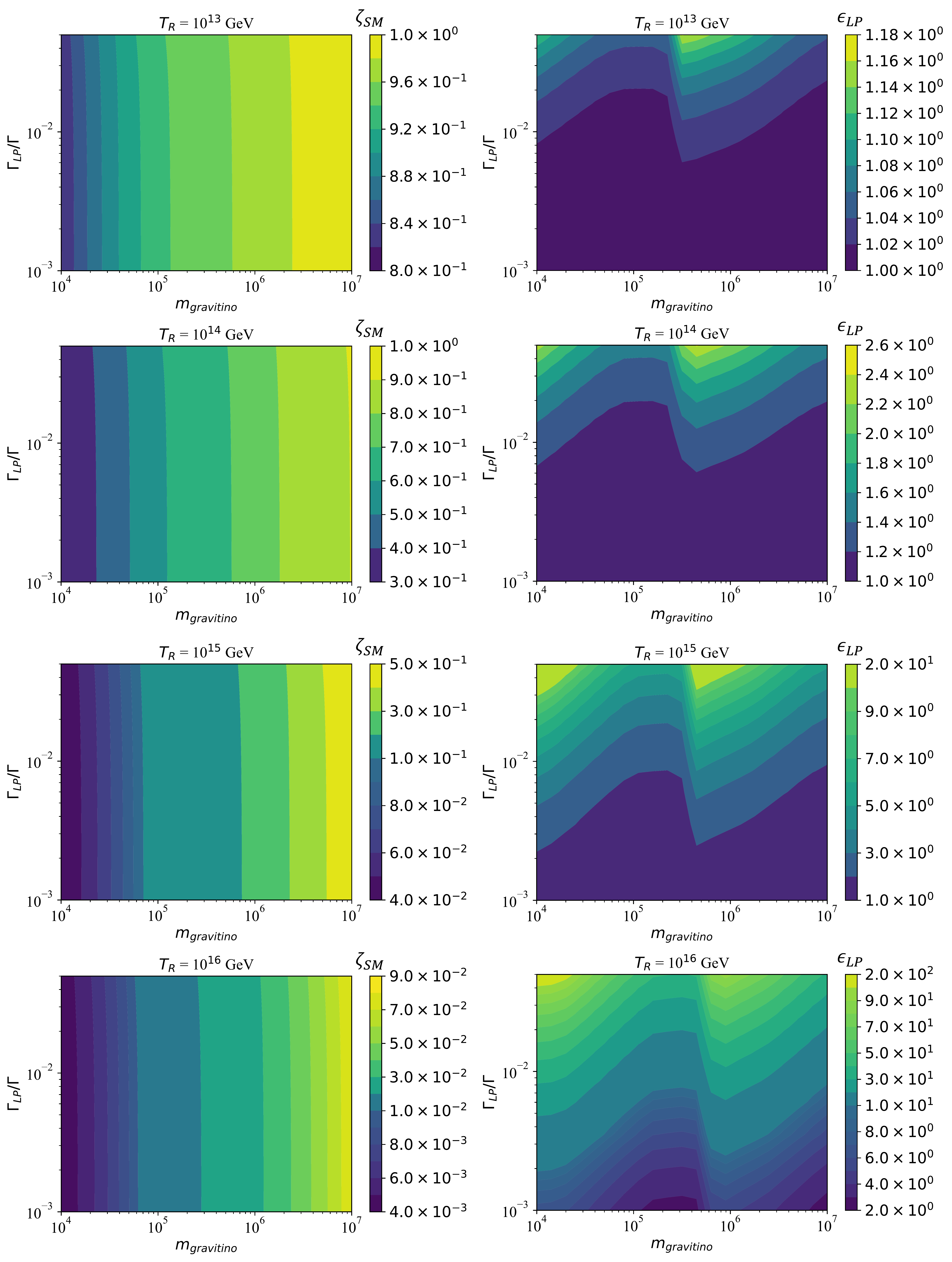}
    \caption{Dilution factor $\zeta_{SM}$ (left column) and enhancement factor $\epsilon_{LP}$ (right column) calculated over a range of gravitino masses and branching ratios $\Gamma_{LP}/\Gamma$, for reheating temperatures $T_R$ between 10$^{13}$-10$^{16}$ GeV. The dominant effect on $\zeta_{SM}$ is the gravitino mass, whereas both the gravitino mass and the branching ratio affect $\epsilon_{LP}$. A feature appears in the plots on the right hand side at a gravitino mass between 10$^5$ and 10$^6$ GeV due to the drop in $g_*^{SM}$ from the gravitino decaying around the QCD phase transition.}
    \label{fig:zeta}
\end{figure}

\begin{figure}[th!]
    \centering
    \includegraphics[width=0.87\columnwidth]{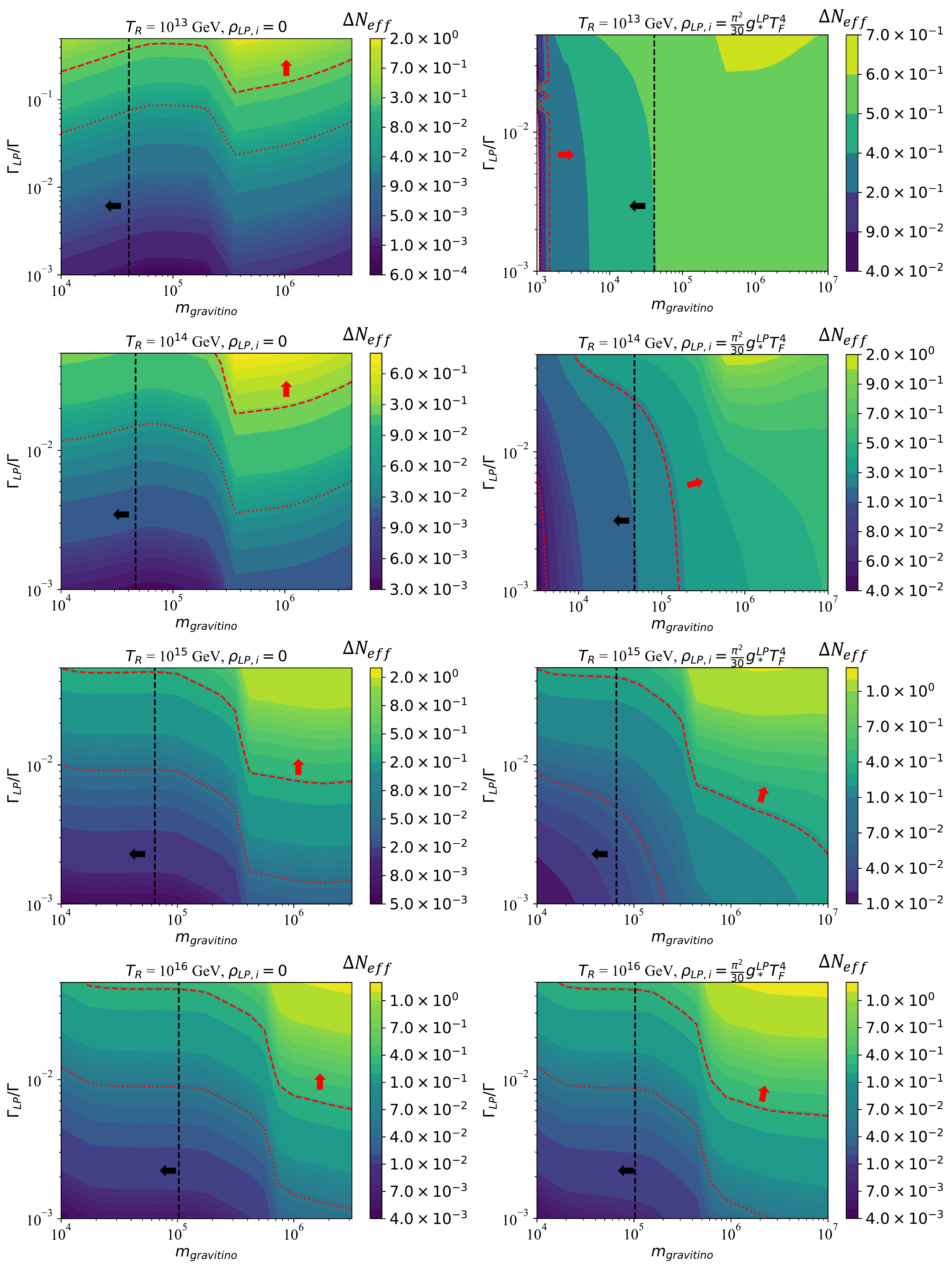}
    \caption{Change in effective number of neutrino species $\Delta N_{eff}$ calculated over a range of gravitino masses and branching ratios $\Gamma_{LP}/\Gamma$, for reheating temperatures $T_R$ between 10$^{13}$-10$^{16}$ GeV, where $\rho_{LP,i} = 0$ (left column) or $\rho_{LP,i} = \frac{\pi}{30}g_*^{LP}T_{F}^4$ (right column). 
    The dashed red line corresponds to the current $2\sigma$ confidence bound on $\Delta N_{eff}$ from Planck's 2018 data \cite{Planck2018}. 
    The dotted red line represents the projected 2$\sigma$ bound on $\Delta N_{eff}$ from CMB-S4 \cite{CMBS4}.
    Gravitino masses to the left of the black dashed line indicate some gravitino decay during or after BBN.
    }
    \label{fig:DeltaNeff}
\end{figure}

We will use the exact numerical method described in the previous section to solve for $\Delta N_{eff}$ while varying over the gravitino mass $m_G$, the reheating temperature $T_R$ and the branching ratio $\Gamma_{LP}/\Gamma$. Hence, we need to determine the decay rate $\Gamma$ of the gravitino, along with the initial cosmological energy densities before the decay, $\rho_{G,i}$, $\rho_{SM,i}$ and $\rho_{LP,i}$; where these are the initial densities of the gravitino, the Standard Model radiation and the light particle respectively. The decay rate of the gravitino is expected to be
\begin{equation}
    \Gamma = \frac{m_G^3}{M_{Pl}^2},
\end{equation}
where $m_G$ $\geq$ 10$^4$ to ensure that the gravitino decays before Big Bang Nucleosynthesis (BBN). This is a condition selected in order to not alter the primordial helium and deuterium abundances that we observe today \cite{Weinberg1982}. The energy density of the gravitino before its decay is characterized in terms of its yield,
\begin{equation}
    \rho_{G,i} = m_G n_{G,i} = m_G s Y_{G, i}.
\label{eq:rho_Gi}
\end{equation}
The yield is defined to be $Y \equiv n/s$, where $n$ is the number density. In this case, the entropy density $s$ of the Universe is dominated by radiation, so the sum of Eq. \ref{eq:s_vs_rho} for the Standard Model and the light particle can be used. The yield of the gravitino at BBN has been determined to be \cite{Pradler2007Electroweak}
\begin{equation}
    Y_{G}(T_{BBN}) = \sum_{\alpha=1}^3 \left( 1 + \frac{M_{\alpha}^2(T_R)}{3m_G^2} \right) y_{\alpha} g_{\alpha}^2(T_R) \text{ ln}\left(\frac{k_{\alpha}}{g_{\alpha}(T_R)}\right) \left( \frac{T_R}{10^{10} \text{ GeV}} \right),
    \label{eq:yg}
\end{equation}
where the subscript $\alpha$ = 1, 2 or 3 corresponds to U(1)$_Y$, SU(2)$_L$ and SU(3)$_c$ respectively, $y_{\alpha}$ = (0.653, 1.604, 4.276) $\times$ 10$^{-12}$, $k_{\alpha}$ = (1.266, 1.312, 1.271), $M_{\alpha}$ = ($M_1$, $M_2$, $M_3$) are the gaugino mass parameters, and $g_{\alpha}$ = ($g'$, $g$, $g_s$) are the gauge couplings, where the latter two parameters are temperature dependent. Here, for simplicity and ease of comparison we take $M_1 = M_2 = M_3 = 10^3$ GeV, and have chosen $g_{\alpha}$ according to \cite{Pradler2007Electroweak}\footnote{In the case that the lightest supersymmetric partner is stabilized by R-parity, having all gaugino masses this light is probably excludable on a number of observational grounds. However, the exact value of $M_\alpha$ does not affect $Y_G$ in the $M_\alpha \ll m_G$ regime in \eqref{eq:yg}, and we note that these values would be permitted in supersymmetric models with R-parity violation.}. We note that our results will be independent of the exact values of $M_{\alpha}$ so long as $M_\alpha \ll m_G$.
Since the universe is radiation-dominated for the parameter space we study, $n_G \propto \rho_G \propto 1/a^3 \propto T^{3}$, and since the entropy density is dominated by radiation, $s \propto T^3$. Therefore, at the time of gravitino decay in our scenario we can approximate
\begin{equation}
    Y_G(T_F) = Y_G(T_{BBN}),
\end{equation}
where $T_F$ is the temperature at which the light particle freezes out, before the gravitino decays, chosen to be 10 times the temperature at which $\Gamma$ = $H$. The energy density of the Standard Model radiation before the gravitino decay is given by
\begin{equation}
    \rho_{SM,i} = \frac{\pi}{30}g_*^{SM}T_{F}^4,
\end{equation}
where $g_*^{SM}$ depends on $T_F$ according to Figure 1 from \cite{Husdal2016}. The energy density of the light particle before the gravitino decay is given by either
\begin{equation}
    \rho_{LP,i} = \frac{\pi}{30}g_*^{LP}T_{F}^4
\end{equation}
in the case where the light particle is already created before the decay by a mechanism independent of the gravitino, or by $\rho_{LP,i} = 0$ in the case where the light particle is only produced as a decay product of the gravitino. For our computations, we have chosen $g_*^{LP} = 1$ as for scalar or pseudoscalar bosons. When $\rho_{LP,i} = 0$, $\epsilon$ = 0, so we cannot use Eq. \ref{eq:DeltaNeff}. Instead, we calculate $\Delta N_{eff}$ directly from the energy densities:
\begin{equation}
    \Delta N_{eff} = \frac{4g}{7} \frac{\rho_{LP}}{\rho_{SM}}.
\end{equation}

In Figure \ref{fig:rhovsa}, the evolution of the energy density of the gravitino ($\rho_G$), of the Standard Model radiation ($\rho_{SM}$) and of the light particle ($\rho_{LP}$) is shown for a gravitino mass $m_G = 2 \times 10^4$ GeV, a branching ratio $\Gamma_{LP}/\Gamma$ = 0.01, and a reheating temperature varying from $T_R = 10^{13}-10^{16}$ GeV. The left column and the right column present the scenario in which $\rho_{LP,i} = 0$ and $\rho_{LP,i} = \frac{\pi}{30}g_*^{LP}T_{F}^4$ respectively. We can observe that for $T_R = 10^{13}$ GeV, the energy density of the gravitino is always smaller than that of the Standard Model, so the decay doesn't result in a significant dilution of the light particle. As the reheating temperature increases, the energy density of the gravitino becomes bigger in comparison to that of the Standard Model just before the decay, resulting in a greater increase in the Standard Model energy density after the decay.

In Figure \ref{fig:zeta}, the values of $\zeta_{SM}$ and $\epsilon_{LP}$ are shown for gravitino masses between 10$^4$-10$^7$ GeV, branching ratios between 0.001-0.05, and reheating temperatures between 10$^{13}$-10$^{16}$ GeV. We can observe that the dominant effect on $\zeta_{SM}$ is the gravitino mass, whereas both the gravitino mass and the branching ratio affect $\epsilon_{LP}$ since there is a bigger change in the entropy being injected into the Standard Model bath in the latter plot. Furthermore, a feature is apparent in the plots on the right hand side at a gravitino mass between 10$^5$ and 10$^6$ GeV. This is due to the drop in $g_*^{SM}$ from the gravitino decaying around the QCD phase transition.

In Figure \ref{fig:DeltaNeff}, the value of $\Delta N_{eff}$ is shown for gravitino masses between 10$^4$-10$^7$ GeV, branching ratios between 0.001-0.05, and reheating temperatures between 10$^{13}$-10$^{16}$ GeV, although some of the axes have been extended to include the bounds described below. The branching ratio seems to be the dominant factor in determining $\Delta N_{eff}$. 

As in the previous figure, a distinct feature is apparent at a gravitino mass between 10$^5$ and 10$^6$ GeV, which is due to the decrease in $g_*^{SM}$ after the QCD phase transition causing $\Delta N_{eff}$ to decrease for lower gravitino masses. At high enough gravitino masses, the decay rate increases such that the gravitino decays before the QCD phase transition. For a given branching ratio, the gravitino decays into more light particle degrees of freedom when $g_*^{SM}$ is large, which is true before the QCD phase transition. Since $\Delta N_{eff}$ increases with the number of light particle degrees of freedom, we expect $\Delta N_{eff}$ to be larger when the gravitino decays before the QCD phase transition, which happens when the gravitino mass is large enough. 

At low reheating temperature, the gravitino mass affects $\Delta N_{eff}$ in two competing ways. First, from Eq. \ref{eq:rho_Gi}, $\rho_{G,i} \propto m_G$. Indeed, $Y$ is effectively independent of $m_G$ in the limit we have chosen $m_G \gg M_{\alpha}$ \eqref{eq:yg}. Therefore, as the gravitino mass increases, the resulting gravitino abundance scales up linearly, and more energy is dumped into the Standard Model bath, causing $\Delta N_{eff}$ to decrease. Second, $\Gamma \propto m_G^3$, so as the gravitino mass gets larger, the gravitino decays earlier, such that its energy density is still small compared to that of the Standard Model (since it will have spent less time redshifting like matter while the SM redshifts like radiation). Therefore, the gravitino cannot dump enough entropy into the Standard Model to dilute $\Delta N_{eff}$, and this parameter increases. However, for high enough reheating temperatures, the energy density of the gravitino will always dominate over that of the Standard Model at the time of its decay, so only the decay rate to the light particle affects $\Delta N_{eff}$, causing $\Delta N_{eff}$ to increase more consistently with mass.

\FloatBarrier
Since the current best estimate for $N_{eff}$ for the Standard Model is 3.043 \cite{Cielo2023}, and the latest Planck measurement has found $N_{eff}$ = 2.99 $\pm$ 0.17 \cite{Planck2018}, the current $2\sigma$ bound on $\Delta N_{eff}$ is 0.287. The projected 2$\sigma$ bound on $\Delta N_{eff}$ from CMB-S4 is 0.055 assuming 1' beams, 1 $\mu$K-arcmin temperature noise and a sky fraction of 0.5 \cite{CMBS4}. These bounds allow us to rule out large branching ratios. Moreover, gravitino masses below a certain threshold depending on the reheating temperature are ruled out because they would imply that the gravitino decays during or after BBN \cite{Weinberg1982}. In the case where $\rho_{LP,i} = \frac{\pi}{30}g_*^{LP}T_{F}^4$, the parameter space is completely ruled out by the combination of these two bounds for $T_R \lesssim 10^{13}$ GeV using the current bounds and for $T_R \lesssim 10^{14}$ GeV using the projected bounds.

\section{Conclusion}
\label{sec:conc}
We have studied how a heavy state decaying into both the Standard Model bath and into a light particle beyond the Standard Model before neutrino decoupling, but after the light particle has frozen out, changes $\Delta N_{eff}$, and found that $\Delta N_{eff}$ will be rescaled by a factor of $\left( \zeta_{SM}\epsilon_{LP} \right)^{4/3}$, where $\zeta_{SM}$ parameterizes the amount the heavy state dilutes non-Standard Model degrees of freedom, and $\epsilon_{LP}$ parameterizes how much it decays into the light particle. If the branching ratio for decay into the light particle is small, the decaying heavy particle dumps entropy mostly into the Standard Model. This causes the energy density of the light particles to become less significant compared to the energy density of the Standard Model, which results in a smaller $\Delta N_{eff}$. Therefore, light particles beyond the Standard Model that would seemingly be ruled out by $N_{eff}$ measurements may still be allowed. Moreover, if a new light particle were found, its effective freeze-out temperature and contribution to $\Delta N_{eff}$, could be different than expected in the heavy state decay scenario.

However, in this case measuring $\Delta N_{eff}$ could be used as evidence for a heavy decaying particle. By examining a more specific model in which the heavy particle is a gravitino, we used current and projected bounds on $N_{eff}$ to place constraints on the gravitino mass and the branching ratio for decay into the light particle for different values of the reheating temperature of the Universe. We have found that $N_{eff}$ limits the decay rate of the gravitino into the light particle, and took into account that forbidding the gravitino from decaying after BBN bounds the gravitino mass from below.

This work also has applications beyond the traditional case of a light, fundamental relic considered here. One such application would be sharpening cosmological predictions for composite dark matter models where bound states are formed through interactions mediated by a light scalar field \cite{Wise:2014jva,Gresham:2017cvl,Acevedo:2020avd}. In this case, the dark matter might annihilate to an effectively massless light scalar, and both this and the composite density may later be depleted by a field decaying \cite{Bramante2017,Acevedo:2020avd,Acevedo:2021kly}, which would result in different predictions for $\Delta N_{eff}$. We look forward to this and other applications of heavy state decay diluted cosmological radiation in future work.

\section*{Acknowledgments}
We are grateful to Aaron Vincent for useful discussions. This work was supported by the Natural Sciences and Engineering
Research Council of Canada (NSERC) and the Arthur B. McDonald Canadian Astroparticle Physics Research Institute. Research at Perimeter Institute is supported in
part by the Government of Canada through the Department of Innovation, Science and Economic Development Canada and by the Province of Ontario through the Ministry of
Colleges and Universities.
\appendix

\section{$\Delta N_{eff}$}
\label{appendix}

We derive the contribution to $N_{eff}$ from any light thermal relic which was in thermal equilibrium with neutrinos at some time before freezing out. First, we must define the energy density $\rho$ and the entropy density $s$ of radiation. The energy density $\rho$ and pressure $p$ of a dilute, weakly-interacting gas of particles in kinetic equilibrium are
\begin{equation}
    \rho = \frac{g}{2\pi^2} \int_{m}^{\infty} \frac{(E^2 - m^2)^{1/2}}{e^{(E-\mu)/T} \pm 1} E^2 \,dE.
\end{equation}
and
\begin{equation}
    p = \frac{g}{6\pi^2} \int_{m}^{\infty} \frac{(E^2 - m^2)^{3/2}}{e^{(E-\mu)/T} \pm 1} E^2 \,dE.
\end{equation}
where the '+' is used for Fermi-Dirac species and the '-' is used for Bose-Einstein species, $g$ is the number of internal degrees of freedom (distinct spin states), $E$ is the internal energy, $m$ is the mass, $\mu$ is the chemical potential and $T$ is the temperature. In the relativistic limit ($T \gg m$) and for $T \gg \mu$, the energy density is
\begin{equation}
    \rho (T) = 
\begin{cases}
    \frac{\pi^2}{30} g T^4 & \text{for bosons}, \\
    \frac{7}{8} \frac{\pi^2}{30} g T^4 & \text{for fermions}
\end{cases}
\end{equation}
and the pressure is
\begin{equation}
    p = \frac{\rho}{3}.
\end{equation}
We can define the entropy density and calculate it in terms of $\rho$:
\begin{equation}
    s \equiv \frac{S}{V} = \frac{\rho + p}{T} = \frac{4\rho}{3T}.
\end{equation}
Therefore, we can write the energy density and the entropy density as
\begin{equation}
\begin{cases}
    \rho(T) = \frac{\pi^2}{30} g_* T^4,  \\
    s(T) = \frac{4}{3} \frac{\pi^2}{30} g_{*s} T^3,
\end{cases}
\end{equation}
where
\begin{equation}
    g_* = \sum_{i = bosons} g_i \left( \frac{T_i}{T_{\gamma}} \right)^4 + \frac{7}{8}\sum_{i = bosons} g_i \left( \frac{T_i}{T_{\gamma}} \right)^4
\label{eq:g*}
\end{equation}
is the number of spin states for all particles and antiparticles, with an extra factor of $\frac{7}{8}$ for fermions, and
\begin{equation}
    g_{*s} = \sum_{i = bosons} g_i \left( \frac{T_i}{T_{\gamma}} \right)^3 + \frac{7}{8}\sum_{i = bosons} g_i \left( \frac{T_i}{T_{\gamma}} \right)^3.
\end{equation}
We can show that $g_* = g_{*s}$ when all relativistic species are in equilibrium at the same temperature.

Next, we will use the fact that the comoving entropy density of the Universe is conserved to write the energy density in radiation of the Universe in terms of the photon temperature and the parameter $N_{eff}$. The statement of comoving entropy density conservation is the following:
\begin{equation}
    a^3 s(T) = a^3g_{*s}T^3 = \text{const},
\label{eq:cons_entropy}
\end{equation}
where $a$ is the scale factor of the Universe. Since both the comoving entropy density of the Standard Model radiation bath (with temperature $T_{\gamma}$), composed of all relativistic particles present in the Universe, and of the neutrino (with temperature $T_{\nu}$) are conserved, noting that $g_{*s}$ is constant for the neutrinos,
\begin{equation}
    a_{before}^3 g_{*s}^{before} T_{\gamma, before}^3 = a_{after}^3 g_{*s}^{after} T_{\gamma, after}^3
\label{eq:entropy_bef_aft}
\end{equation}
and
\begin{equation}
    a_{before}^3 T_{\nu, before}^3 = a_{after}^3 T_{\nu, after}^3.
\label{eq:entropy_bef_aft_nu}
\end{equation}
Dividing Eq. \ref{eq:entropy_bef_aft} by Eq. \ref{eq:entropy_bef_aft_nu}, we find
\begin{equation}
\frac{g_*^{before} T_{\gamma, before}^3}{T_{\nu, before}^3} = \frac{g_*^{after} T_{\gamma, after}^3}{T_{\nu, after}^3}.
\end{equation}
After neutrino decoupling, but before electron-positron annihilation, $g_*^{before} = 2 + \frac{7}{8}(2+2) = \frac{11}{2}$, where the bosons are photons and the fermions are electrons and positrons. After electron-positron annihilation, we are left with only photon degrees of freedom, so $g_*^{after} = 2$. In the instantaneous neutrino decoupling limit,
\begin{equation}
     T_{\gamma, before} = T_{\nu, before} \implies \frac{T_{\nu, after}}{T_{\gamma, after}} = \left( \frac{g_*^{after}}{g_*^{before}} \right)^{1/3} = \left( \frac{4}{11} \right)^{1/3},
\label{eq:before_after}
\end{equation}
which means that the annihilation of electrons and positrons raised the temperature of photons relative to that of neutrinos by a factor of $\left( \frac{11}{4} \right)^{1/3}$. After electron-positron annihilation, assuming that there exist three neutrinos and three antineutrinos with one spin state each, the radiation density of the Universe is given by
\begin{equation}
     \rho_r = \frac{\pi^2}{30} \left[ 2T_{\gamma}^4 + 6\frac{7}{8}T_{\nu}^4 \right] = \frac{\pi^2}{15} \left[ 1 + 3\frac{7}{8} \left( \frac{4}{11} \right)^{4/3} \right] T_{\gamma}^4.
\end{equation}
The factor of 3 in the second term is the effective number of neutrino species $N_{eff}$. We can parameterize $\rho_r$ by $N_{eff}$:
\begin{equation}
    \rho_r = \frac{\pi^2}{15} \left[1 + \frac{7}{8} \left(\frac{4}{11} \right)^{4/3} N_{eff} \right]T_{\gamma}^4.
\end{equation}
As described above, $N_{eff} = 3$ in the instantaneous neutrino decoupling approximation. In the real Universe, the current best estimate is $N_{eff} = 3.043$ \cite{Cielo2023} due to the incomplete decoupling of neutrinos during $e^- - e^+$ annihilation and QED plasma effects.

Finally, we will calculate the expected contribution to $N_{eff}$ from any light particle (LP) which was in thermal equilibrium with neutrinos at some time before freezing out and which freezes out before the neutrinos decouple from the photon bath. Before neutrino decoupling, given conservation of comoving entropy density for both the light particle (with temperature $T_{LP}$), whose value of $g_{*s}$ is constant, and for the Standard Model radiation bath (with temperature $T_{\nu}$),
\begin{equation}
    \left( \frac{T_{LP}}{T_{\nu}} \right)^3 = \frac{g_*(T_{\nu-decoupling})}{g_*(T_F)} = \frac{43/4}{g_*(T_F)},
\label{eq:T_LP_nu}
\end{equation}
where $T_{\nu-decoupling}$ is the temperature before neutrino decoupling and $T_F$ is the temperature at which the light particle decouples (freezes out) from the neutrino bath. The relativistic bosons are photons and the relativistic fermions are electrons, positrons, neutrinos and antineutrinos, so $g_*(T_{before}) = 2 + \frac{7}{8}(4 + 6) = \frac{43}{4}$. We can now calculate the energy density in radiation of the Universe. If the light particle is a boson,
\begin{equation}
\begin{split}
     \rho_r & = \frac{\pi^2}{30} \left[ 2T_{\gamma}^4 + 6\frac{7}{8}T_{\nu}^4 + gT_{LP}^4 \right] \\ & = \frac{\pi^2}{15} \left[ 1 + \frac{7}{8} \left( \frac{T_{\nu}}{T_{\gamma}} \right)^4 \left(\frac{8}{7} \frac{g}{2} \left( 3 + \frac{T_{relic}}{T_{\nu}} \right)^4 \right) \right] T_{\gamma}^4 \\ & = \frac{\pi^2}{15} \left[ 1 + \frac{7}{8} \left( \frac{4}{11} \right)^{4/3} \left(\frac{4g}{7} \left( 3 + \frac{43/4}{g_*(T_F)} \right)^{4/3} \right) \right] T_{\gamma}^4 \\ & \implies \Delta N_{eff} = \frac{4g}{7} \left( \frac{43/4}{g_*(T_F)} \right)^{4/3}
\end{split}
\end{equation}
If the light particle is a fermion,
\begin{equation}
\begin{split}
     \rho_r & = \frac{\pi^2}{30} \left[ 2T_{\gamma}^4 + 6\frac{7}{8}T_{\nu}^4 + g\frac{7}{8}T_{relic}^4 \right] \\ & = \frac{\pi^2}{15} \left[ 1 + \frac{7}{8} \left( \frac{T_{\nu}}{T_{\gamma}} \right)^4 \left(3 + \frac{g}{2} \left( \frac{T_{relic}}{T_{\nu}} \right)^4 \right) \right] T_{\gamma}^4 \\ & = \frac{\pi^2}{15} \left[ 1 + \frac{7}{8} \left( \frac{4}{11} \right)^{4/3} \left(3 + \frac{g}{2} \left( \frac{43/4}{g_*(T_F)} \right)^{4/3} \right) \right] T_{\gamma}^4 \\ & \implies \Delta N_{eff} = \frac{g}{2} \left( \frac{43/4}{g_*(T_F)} \right)^{4/3}
\end{split}
\end{equation}
To summarize,
\begin{equation}
    \Delta N_{eff} = 
\begin{cases}
    \frac{4g}{7} \left(\frac{43/4}{g_*(T_F)} \right)^{4/3} & \text{for a boson}, \\
    \frac{g}{2} \left(\frac{43/4}{g_*(T_F)} \right)^{4/3} & \text{for a fermion},
\end{cases}
\end{equation}
which is Eq.~\eqref{eq:N_eff}.

We can also note that the contribution to $\Delta N_{eff}$ from a non-relativistic particle is negligible. In this case, the energy density in radiation in terms of the photon temperature is
\begin{equation}
    \rho_r = T_{\gamma}^4 \sum_{i} \left( \frac{T_i}{T_{\gamma}} \right)^4 \frac{g_i}{2\pi^2} \int_{x_i}^\infty \frac{(u^2 - x_i^2)^{1/2} u^2 du}{e^{u-y_i} \pm 1},
\end{equation}
where $x_i = m_i/T_{\gamma}$ and $y_i = \mu_i/T_{\gamma}$. Since $m \gg T$ for a non-relativistic species, its energy density in radiation is exponentially suppressed.

\bibliographystyle{JHEP.bst}
\bibliography{references}

\providecommand{\href}[2]{#2}\begingroup\raggedright\begin{thebibliography}{10}

\bibitem{rotation_curves}
V.~C. {Rubin} and J.~{Ford}, W.~Kent, \emph{{Rotation of the Andromeda Nebula
  from a Spectroscopic Survey of Emission Regions}},
  \href{http://dx.doi.org/10.1086/150317}{\emph{The Astrophysical Journal} {\bf
  159} (Feb., 1970) 379}.

\bibitem{Rubin:1980zd}
V.~C. Rubin, N.~Thonnard and W.~K. Ford, Jr., \emph{{Rotational properties of
  21 SC galaxies with a large range of luminosities and radii, from NGC 4605 /R
  = 4kpc/ to UGC 2885 /R = 122 kpc/}},
  \href{http://dx.doi.org/10.1086/158003}{\emph{Astrophys. J.} {\bf 238} (1980)
  471}.

\bibitem{bullet_cluster}
D.~Clowe, M.~Bradač, A.~H. Gonzalez, M.~Markevitch, S.~W. Randall, C.~Jones
  et~al., \emph{A direct empirical proof of the existence of dark matter*},
  \href{http://dx.doi.org/10.1086/508162}{\emph{The Astrophysical Journal} {\bf
  648} (aug, 2006) L109}.

\bibitem{CMB_power_spectra}
{Planck Collaboration}, {Aghanim, N.}, {Akrami, Y.}, {Ashdown, M.}, {Aumont,
  J.}, {Baccigalupi, C.} et~al., \emph{Planck 2018 results - v. cmb power
  spectra and likelihoods},
  \href{http://dx.doi.org/10.1051/0004-6361/201936386}{\emph{A\&A} {\bf 641}
  (2020) A5}.

\bibitem{Dine:1982ah}
M.~Dine and W.~Fischler, \emph{{The Not So Harmless Axion}},
  \href{http://dx.doi.org/10.1016/0370-2693(83)90639-1}{\emph{Phys. Lett. B}
  {\bf 120} (1983) 137--141}.

\bibitem{Preskill:1982cy}
J.~Preskill, M.~B. Wise and F.~Wilczek, \emph{{Cosmology of the Invisible
  Axion}}, \href{http://dx.doi.org/10.1016/0370-2693(83)90637-8}{\emph{Phys.
  Lett. B} {\bf 120} (1983) 127--132}.

\bibitem{Holdom:1985ag}
B.~Holdom, \emph{{Two U(1)'s and Epsilon Charge Shifts}},
  \href{http://dx.doi.org/10.1016/0370-2693(86)91377-8}{\emph{Phys. Lett. B}
  {\bf 166} (1986) 196--198}.

\bibitem{Masso:1995tw}
E.~Masso and R.~Toldra, \emph{{On a light spinless particle coupled to
  photons}}, \href{http://dx.doi.org/10.1103/PhysRevD.52.1755}{\emph{Phys. Rev.
  D} {\bf 52} (1995) 1755--1763},
  [\href{http://arxiv.org/abs/hep-ph/9503293}{{\tt hep-ph/9503293}}].

\bibitem{Hu:2000ke}
W.~Hu, R.~Barkana and A.~Gruzinov, \emph{{Cold and fuzzy dark matter}},
  \href{http://dx.doi.org/10.1103/PhysRevLett.85.1158}{\emph{Phys. Rev. Lett.}
  {\bf 85} (2000) 1158--1161},
  [\href{http://arxiv.org/abs/astro-ph/0003365}{{\tt astro-ph/0003365}}].

\bibitem{McDonald:2001vt}
J.~McDonald, \emph{{Thermally generated gauge singlet scalars as
  selfinteracting dark matter}},
  \href{http://dx.doi.org/10.1103/PhysRevLett.88.091304}{\emph{Phys. Rev.
  Lett.} {\bf 88} (2002) 091304},
  [\href{http://arxiv.org/abs/hep-ph/0106249}{{\tt hep-ph/0106249}}].

\bibitem{Boehm:2003hm}
C.~Boehm and P.~Fayet, \emph{{Scalar dark matter candidates}},
  \href{http://dx.doi.org/10.1016/j.nuclphysb.2004.01.015}{\emph{Nucl. Phys. B}
  {\bf 683} (2004) 219--263}, [\href{http://arxiv.org/abs/hep-ph/0305261}{{\tt
  hep-ph/0305261}}].

\bibitem{Feng:2008ya}
J.~L. Feng and J.~Kumar, \emph{{The WIMPless Miracle: Dark-Matter Particles
  without Weak-Scale Masses or Weak Interactions}},
  \href{http://dx.doi.org/10.1103/PhysRevLett.101.231301}{\emph{Phys. Rev.
  Lett.} {\bf 101} (2008) 231301}, [\href{http://arxiv.org/abs/0803.4196}{{\tt
  0803.4196}}].

\bibitem{Feng:2008mu}
J.~L. Feng, H.~Tu and H.-B. Yu, \emph{{Thermal Relics in Hidden Sectors}},
  \href{http://dx.doi.org/10.1088/1475-7516/2008/10/043}{\emph{JCAP} {\bf 10}
  (2008) 043}, [\href{http://arxiv.org/abs/0808.2318}{{\tt 0808.2318}}].

\bibitem{Hooper:2008im}
D.~Hooper and K.~M. Zurek, \emph{{A Natural Supersymmetric Model with MeV Dark
  Matter}}, \href{http://dx.doi.org/10.1103/PhysRevD.77.087302}{\emph{Phys.
  Rev. D} {\bf 77} (2008) 087302}, [\href{http://arxiv.org/abs/0801.3686}{{\tt
  0801.3686}}].

\bibitem{Kaplinghat:2013yxa}
M.~Kaplinghat, S.~Tulin and H.-B. Yu, \emph{{Direct Detection Portals for
  Self-interacting Dark Matter}},
  \href{http://dx.doi.org/10.1103/PhysRevD.89.035009}{\emph{Phys. Rev. D} {\bf
  89} (2014) 035009}, [\href{http://arxiv.org/abs/1310.7945}{{\tt 1310.7945}}].

\bibitem{Boddy:2014yra}
K.~K. Boddy, J.~L. Feng, M.~Kaplinghat and T.~M.~P. Tait,
  \emph{{Self-Interacting Dark Matter from a Non-Abelian Hidden Sector}},
  \href{http://dx.doi.org/10.1103/PhysRevD.89.115017}{\emph{Phys. Rev. D} {\bf
  89} (2014) 115017}, [\href{http://arxiv.org/abs/1402.3629}{{\tt 1402.3629}}].

\bibitem{Knapen:2017xzo}
S.~Knapen, T.~Lin and K.~M. Zurek, \emph{{Light Dark Matter: Models and
  Constraints}},
  \href{http://dx.doi.org/10.1103/PhysRevD.96.115021}{\emph{Phys. Rev. D} {\bf
  96} (2017) 115021}, [\href{http://arxiv.org/abs/1709.07882}{{\tt
  1709.07882}}].

\bibitem{Boehm:2012gr}
C.~Boehm, M.~J. Dolan and C.~McCabe, \emph{{Increasing Neff with particles in
  thermal equilibrium with neutrinos}},
  \href{http://dx.doi.org/10.1088/1475-7516/2012/12/027}{\emph{JCAP} {\bf 12}
  (2012) 027}, [\href{http://arxiv.org/abs/1207.0497}{{\tt 1207.0497}}].

\bibitem{Boehm:2013jpa}
C.~Boehm, M.~J. Dolan and C.~McCabe, \emph{{A Lower Bound on the Mass of Cold
  Thermal Dark Matter from Planck}},
  \href{http://dx.doi.org/10.1088/1475-7516/2013/08/041}{\emph{JCAP} {\bf 08}
  (2013) 041}, [\href{http://arxiv.org/abs/1303.6270}{{\tt 1303.6270}}].

\bibitem{Sabti:2019mhn}
N.~Sabti, J.~Alvey, M.~Escudero, M.~Fairbairn and D.~Blas, \emph{{Refined
  Bounds on MeV-scale Thermal Dark Sectors from BBN and the CMB}},
  \href{http://dx.doi.org/10.1088/1475-7516/2020/01/004}{\emph{JCAP} {\bf 01}
  (2020) 004}, [\href{http://arxiv.org/abs/1910.01649}{{\tt 1910.01649}}].

\bibitem{Chu:2022xuh}
X.~Chu, J.-L. Kuo and J.~Pradler, \emph{{Toward a full description of MeV dark
  matter decoupling: A self-consistent determination of relic abundance and
  Neff}}, \href{http://dx.doi.org/10.1103/PhysRevD.106.055022}{\emph{Phys. Rev.
  D} {\bf 106} (2022) 055022}, [\href{http://arxiv.org/abs/2205.05714}{{\tt
  2205.05714}}].

\bibitem{CMBS4}
K.~N. Abazajian, P.~Adshead, Z.~Ahmed, S.~W. Allen, D.~Alonso, K.~S. Arnold
  et~al., \emph{{CMB-S4 Science Book, First Edition}},  2016.
\newblock 10.48550/ARXIV.1610.02743.

\bibitem{SimonsObservatory}
P.~Ade, J.~Aguirre, Z.~Ahmed, S.~Aiola, A.~Ali, D.~Alonso et~al., \emph{The
  simons observatory: science goals and forecasts},
  \href{http://dx.doi.org/10.1088/1475-7516/2019/02/056}{\emph{Journal of
  Cosmology and Astroparticle Physics} {\bf 2019} (feb, 2019) 056}.

\bibitem{Affleck:1984fy}
I.~Affleck and M.~Dine, \emph{{A New Mechanism for Baryogenesis}},
  \href{http://dx.doi.org/10.1016/0550-3213(85)90021-5}{\emph{Nucl. Phys. B}
  {\bf 249} (1985) 361--380}.

\bibitem{Banks:1993en}
T.~Banks, D.~B. Kaplan and A.~E. Nelson, \emph{{Cosmological implications of
  dynamical supersymmetry breaking}},
  \href{http://dx.doi.org/10.1103/PhysRevD.49.779}{\emph{Phys. Rev. D} {\bf 49}
  (1994) 779--787}, [\href{http://arxiv.org/abs/hep-ph/9308292}{{\tt
  hep-ph/9308292}}].

\bibitem{deCarlos:1993wie}
B.~de~Carlos, J.~A. Casas, F.~Quevedo and E.~Roulet, \emph{{Model independent
  properties and cosmological implications of the dilaton and moduli sectors of
  4-d strings}},
  \href{http://dx.doi.org/10.1016/0370-2693(93)91538-X}{\emph{Phys. Lett. B}
  {\bf 318} (1993) 447--456}, [\href{http://arxiv.org/abs/hep-ph/9308325}{{\tt
  hep-ph/9308325}}].

\bibitem{Moroi:1999zb}
T.~Moroi and L.~Randall, \emph{{Wino cold dark matter from anomaly mediated
  SUSY breaking}},
  \href{http://dx.doi.org/10.1016/S0550-3213(99)00748-8}{\emph{Nucl. Phys. B}
  {\bf 570} (2000) 455--472}, [\href{http://arxiv.org/abs/hep-ph/9906527}{{\tt
  hep-ph/9906527}}].

\bibitem{Randall2016}
L.~Randall, J.~Scholtz and J.~Unwin, \emph{Flooded dark matter and s level
  rise}, \href{http://dx.doi.org/10.1007/jhep03(2016)011}{\emph{Journal of High
  Energy Physics} {\bf 2016} (mar, 2016) }.

\bibitem{Weinberg:1982zq}
S.~Weinberg, \emph{{Cosmological Constraints on the Scale of Supersymmetry
  Breaking}}, \href{http://dx.doi.org/10.1103/PhysRevLett.48.1303}{\emph{Phys.
  Rev. Lett.} {\bf 48} (1982) 1303}.

\bibitem{Dine:1995kz}
M.~Dine, L.~Randall and S.~D. Thomas, \emph{{Baryogenesis from flat directions
  of the supersymmetric standard model}},
  \href{http://dx.doi.org/10.1016/0550-3213(95)00538-2}{\emph{Nucl. Phys. B}
  {\bf 458} (1996) 291--326}, [\href{http://arxiv.org/abs/hep-ph/9507453}{{\tt
  hep-ph/9507453}}].

\bibitem{KolbTurner}
E.~W. Kolb and M.~S. Turner, \emph{{The Early Universe}}, vol.~69.
\newblock 1990,
  \href{http://dx.doi.org/10.1201/9780429492860}{10.1201/9780429492860}.

\bibitem{Planck2018}
{\scshape Planck} collaboration, N.~Aghanim et~al., \emph{{Planck 2018 results.
  VI. Cosmological parameters}},
  \href{http://dx.doi.org/10.1051/0004-6361/201833910}{\emph{Astron.
  Astrophys.} {\bf 641} (2020) A6},
  [\href{http://arxiv.org/abs/1807.06209}{{\tt 1807.06209}}].

\bibitem{Bramante2017}
J.~Bramante and J.~Unwin, \emph{Superheavy thermal dark matter and primordial
  asymmetries}, \href{http://dx.doi.org/10.1007/jhep02(2017)119}{\emph{Journal
  of High Energy Physics} {\bf 2017} (feb, 2017) }.

\bibitem{Martin:1997ns}
S.~P. Martin, \emph{{A Supersymmetry primer}},
  \href{http://dx.doi.org/10.1142/9789812839657_0001}{\emph{Adv. Ser. Direct.
  High Energy Phys.} {\bf 18} (1998) 1--98},
  [\href{http://arxiv.org/abs/hep-ph/9709356}{{\tt hep-ph/9709356}}].

\bibitem{Nastase:2011aa}
H.~Nastase, \emph{{Introduction to Supergravity}},
  \href{http://arxiv.org/abs/1112.3502}{{\tt 1112.3502}}.

\bibitem{Pagels:1981ke}
H.~Pagels and J.~R. Primack, \emph{{Supersymmetry, Cosmology and New TeV
  Physics}}, \href{http://dx.doi.org/10.1103/PhysRevLett.48.223}{\emph{Phys.
  Rev. Lett.} {\bf 48} (1982) 223}.

\bibitem{Weinberg1982}
S.~Weinberg, \emph{Cosmological constraints on the scale of supersymmetry
  breaking}, \href{http://dx.doi.org/10.1103/PhysRevLett.48.1303}{\emph{Phys.
  Rev. Lett.} {\bf 48} (May, 1982) 1303--1306}.

\bibitem{Pradler2007Electroweak}
J.~Pradler, \emph{Electroweak contributions to thermal gravitino production},
  \href{http://arxiv.org/abs/0708.2786}{{\tt 0708.2786}}.

\bibitem{Husdal2016}
L.~Husdal, \emph{On effective degrees of freedom in the early universe},
  \href{http://dx.doi.org/10.3390/galaxies4040078}{\emph{Galaxies} {\bf 4}
  (dec, 2016) 78}.

\bibitem{Cielo2023}
M.~Cielo, M.~Escudero, G.~Mangano and O.~Pisanti, \emph{Neff in the standard
  model at nlo is 3.043},  2023.

\bibitem{Wise:2014jva}
M.~B. Wise and Y.~Zhang, \emph{{Stable Bound States of Asymmetric Dark
  Matter}}, \href{http://dx.doi.org/10.1103/PhysRevD.90.055030}{\emph{Phys.
  Rev. D} {\bf 90} (2014) 055030}, [\href{http://arxiv.org/abs/1407.4121}{{\tt
  1407.4121}}].

\bibitem{Gresham:2017cvl}
M.~I. Gresham, H.~K. Lou and K.~M. Zurek, \emph{{Early Universe synthesis of
  asymmetric dark matter nuggets}},
  \href{http://dx.doi.org/10.1103/PhysRevD.97.036003}{\emph{Phys. Rev. D} {\bf
  97} (2018) 036003}, [\href{http://arxiv.org/abs/1707.02316}{{\tt
  1707.02316}}].

\bibitem{Acevedo:2020avd}
J.~F. Acevedo, J.~Bramante and A.~Goodman, \emph{{Nuclear fusion inside dark
  matter}}, \href{http://dx.doi.org/10.1103/PhysRevD.103.123022}{\emph{Phys.
  Rev. D} {\bf 103} (2021) 123022},
  [\href{http://arxiv.org/abs/2012.10998}{{\tt 2012.10998}}].

\bibitem{Acevedo:2021kly}
J.~F. Acevedo, J.~Bramante and A.~Goodman, \emph{{Accelerating composite dark
  matter discovery with nuclear recoils and the Migdal effect}},
  \href{http://dx.doi.org/10.1103/PhysRevD.105.023012}{\emph{Phys. Rev. D} {\bf
  105} (2022) 023012}, [\href{http://arxiv.org/abs/2108.10889}{{\tt
  2108.10889}}].

\end{thebibliography}\endgroup

\end{document}